\documentclass[preprint,amsmath,amssymb,aps]{revtex4-2}

\usepackage{graphicx}
\usepackage{dcolumn}
\usepackage{bm}
\usepackage{color}
\usepackage{xcolor}
\usepackage[normalem]{ulem}
\usepackage{setspace}

\usepackage[mathlines]{lineno}

\begin{document}

\title{
From creep to flow: Granular materials under cyclic shear\\
}

\author{Ye Yuan$^1$}%
\author{Zhikun Zeng$^1$}%
\author{Yi Xing$^1$}%
\author{Houfei Yuan$^1$}%
\author{Shuyang Zhang$^1$}%

\author{Walter Kob$^{2}$}%
 \email{walter.kob@umontpellier.fr}
 
\author{Yujie Wang$^{1,3,4}$}%
 \email{yujiewang@sjtu.edu.cn}
\affiliation{%
$^1$ School of Physics and Astronomy, Shanghai Jiao Tong University, Shanghai 200240, China\\
$^2$ Laboratoire Charles Coulomb, University of Montpellier and CNRS, 34095 Montpellier, France\\
$^3$ State Key Laboratory of Geohazard Prevention and Geoenvironment Protection, Chengdu University of Technology, Chengdu 610059, China\\
$^4$ Department of Physics, College of Mathematics and Physics, Chengdu University of Technology, Chengdu 610059, China
}%

\date{\today}

\setstretch{1.2}

\begin{abstract}
Granular materials such as sand, powders, and grains are omnipresent
in daily life, industrial applications, and earth-science~\cite{jaeger1996granular}. When unperturbed, they form stable structures that resemble the ones of other amorphous solids like metallic and colloidal glasses~\cite{binder2011glassy}. It is commonly conjectured that all these amorphous materials show a universal mechanical response when sheared slowly, i.e., to have an elastic regime, followed by yielding~\cite{nicolas2018deformation}. 
Here we use X-ray tomography to determine the microscopic dynamics of a cyclically sheared granular system in three dimensions. Independent of the shear amplitude $\Gamma$, the sample shows a cross-over from creep to diffusive dynamics, indicating that granular materials have no elastic response and always yield, in stark contrast to other glasses.  
The overlap function~\cite{jaiswal2016mechanical} reveals that at large $\Gamma$ yielding is a simple cross-over phenomenon, while for small $\Gamma$ it shows features of a first order transition with a critical point at $\Gamma\approx 0.1$ at which one finds a pronounced slowing down and dynamical heterogeneity. Our findings are directly related to the surface roughness of granular particles which induces a micro-corrugation to the potential energy landscape, thus creating relaxation channels that are absent in simple glasses. These processes must be understood for reaching an understanding of the complex relaxation dynamics of granular systems.
\end{abstract}

\maketitle

{\bf Introduction}

Yielding of amorphous materials is ubiquitous, governing a wide range of phenomena like mechanical failure of metallic glasses~\cite{wang2004bulk}, complex rheologies of soft glasses~\cite{bonn2017yield}, or geophysical catastrophes~\cite{ancey2007plasticity}. A typical amorphous solid under quasistatic simple shear will deform elastically at small strain and flow plastically beyond the yielding strain. Depending on whether the material is brittle or ductile, the stress-strain curve shows a drop or a smooth cross-over to a plateau, respectively. 
On the particle level, yielding is a cooperative phenomenon of local plastic events~\cite{schall2007structural,ghosh2017direct}, and computer simulations hint that this phenomenon shares many aspects with a first order phase transition
~\cite{jaiswal2016mechanical,parisi2017shear,ozawa2018random}. However, for real systems the precise nature of this transition has not yet been clarified since such experiments are very challenging.

In contrast to standard glasses, for granular systems there is at present no satisfactory understanding of yielding on the level of the particles. 
It has been suggested that sheared granular solids are marginally stable, i.e., contacts between particles will change irreversibly even under a tiny applied strain, which implies that such systems have no elastic regime~\cite{cates1998jamming}. This view seems to be at odds with experimental results which found that the stress-strain curve of granular solids under simple shear does in fact resemble the one for a glass~\cite{murphy2019transforming}. One possibility to resolve this discrepancy is to consider a different type of driving, i.e., cyclic shear, since it allows to probe directly the reversibility of the particle trajectories and hence the presence of elasticity in the system. 
A number of computational studies have in fact used this setup to investigate the particle dynamics in soft-sphere jammed states and model glasses~\cite{royer2015precisely,kawasaki2016macroscopic,dagois2017softening,nagasawa2019classification,das2020unified,priezjev2013heterogeneous,fiocco2013oscillatory,regev2015reversibility,leishangthem2017yielding,jin2018stability,yeh2020glass}. These works, as well as related experiments on soft glasses~\cite{keim2014mechanical,knowlton2014microscopic,nagamanasa2014experimental}, have indeed revealed a reversible-irreversible transition as the cyclic shear amplitude $\Gamma$ increases, thus supporting the existence of an elastic behavior if $\Gamma \lesssim 0.1$.  However, these findings are in contrast with experimental results for cyclically sheared granular particles in three dimensions (3D) which indicate the absence of an elastic regime, since the particle trajectories are found to be irreversible, at least for the values of $\Gamma$ considered, i.e., $0.07 \leq \Gamma \leq 0.26$~\cite{kou2017granular,kou2018translational}.

Although a few experimental studies have probed the microscopic dynamics of driven 3D granular materials~\cite{pouliquen2003fluctuating,slotterback2012onset,denisov2016universality}, none of them has made an explicit connection between this dynamics and the phenomenon of yielding. Establishing this link is important not only for understanding the plasticity of amorphous materials but also for developing reliable granular constitution laws~\cite{forterre2008flows}. The goal of the present work is therefore to identify this connection.\\

{\bf Cyclic shear experiment}

In this work, we experimentally investigate a cyclically sheared granular system using a X-ray tomography technique~\cite{kou2017granular,kou2018translational, li2021microscopic,yuan2021experimental}. The system contains $\approx$14000 $50:50$ bidisperse plastic spherical beads of diameters 5~mm and 6~mm and no crystallization was detected. Particles are placed in a shear cell and cyclically sheared with an amplitude $\Gamma$~\cite{kou2017granular,xing2021x}. The size of the shear cell at shear $\gamma = 0$ is $24d\times24d\times24d$, where $d$ is the diameter of the small beads, and which in the following will be taken as the unit length. The shear rate is small so that we are in quasistatic conditions with an inertia number less than $10^{-3}$~\cite{gdr2004dense}. 

Beads are initially placed in the cell, forming a reproducible loose packings, and then compacted by cyclic shear until the steady state is reached (see Extended Data Fig. 1). X-ray tomography scans are taken at $\gamma=0$, with a periodicity between 1 and 10 cycles, from which we extract the microscopic structure of the system and the dynamics of the particles. To improve the statistics of the results, we use $3 \sim 5$ independent realizations for each $\Gamma$, and to mitigate finite size effects we exclude the particles located closer than $6d$ from the boundaries.\\ 

{\bf Steady-state dynamics}

\begin{figure}[t!]
\centering
\includegraphics[width=12cm]{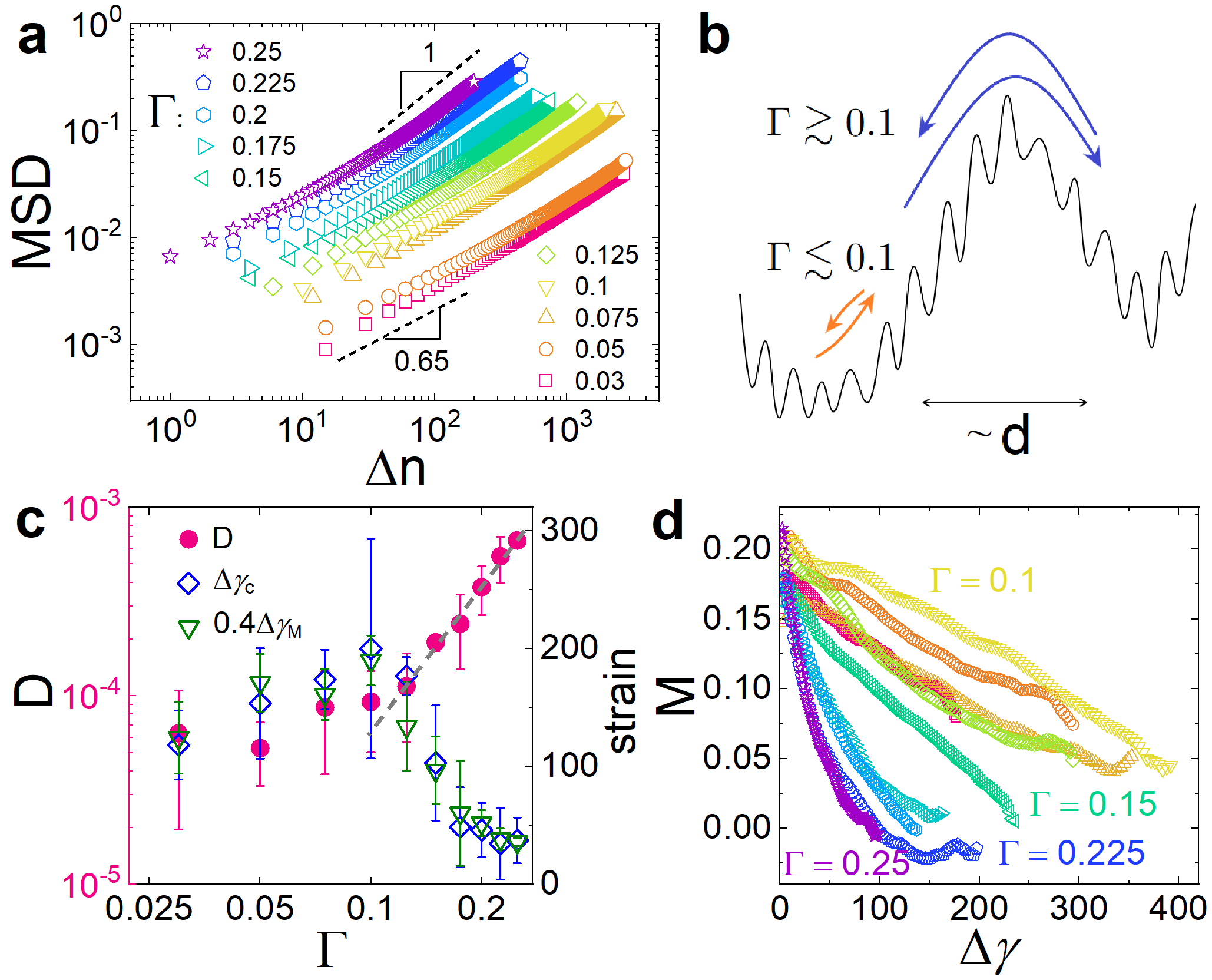}
\caption{
{\bf Bulk dynamics of cyclic shear for different shear amplitudes $\Gamma$.} (a) Mean squared displacement in horizontal directions vs. shear cycle number $\Delta n$. A universal crossover from sub-diffusion MSD $\propto \Delta n^{0.65}$ to normal diffusion (dashed lines) is observed, corresponding to the $\Gamma$-dependent yielding point. (b) Schematics of how a granular system explores its potential energy landscape. Due to the particle surface roughness, the PEL has not only a structure on the particle scale $d$, but also a micro-corrugation. Starting in the left metabasin of the PEL, the double arrows indicate the back and forth motions of the system during a shear cycle. Depending on $\Gamma$, the system does/(does not) leave the MB during a cycle (blue and orange arrows, respectively). Yielding means that the system has overcome permanently the PEL barrier (on the scale $d$). (c) $\Gamma$-dependence of the diffusion coefficient $D$ (circles), the yielding strain $\Delta\gamma_c$ (diamonds), and the strain $\Delta\gamma_M$ at which memory is lost (triangles). Note that $\Delta\gamma_c \approx 0.4\Delta\gamma_M$ for all $\Gamma$. Error bars represent the standard deviations from $3-5$ independent realizations. (d) Memory $M$, defined in Eq.~(\ref{memof}), as a function of $\Delta \gamma$, from which $\Delta\gamma_M$ is obtained by $M(\Delta\gamma_M) = 0$. For $\Gamma \leq 0.175$, $\Delta \gamma_M$ is estimated from a linear extrapolation of $M(\Delta \gamma)$. Color codes are the same as in Fig.~\ref{fig_1_msd}(a).\\
}
\label{fig_1_msd}
\end{figure}

For small $\Gamma$ the dynamics of the particles is basically isotropic although weak convection caused by gravity emerges for large $\Gamma$ (see Extended Data Fig.~2). Hence we focus here on the dynamics in the horizontal $x$-and $y$-directions. Figure~\ref{fig_1_msd}(a) presents the mean squared displacement (MSD), $\langle \delta h^2(\Delta n)\rangle$, as a function of the number of cycles $\Delta n$, where $\delta h$ is the horizontal displacement and $\langle .\rangle$ stands for the average over different particles, starting configurations, and realizations. At large $\Delta n$ we find the expected diffusive growth while at small $\Delta n$ the MSD shows a power-law dependence, with a $\Gamma$-independent exponent $\approx$\;0.65. Such a universal creep dynamics at small $\Delta n$ demonstrates that our system has no caging or elastic regime, in agreement with earlier findings~\cite{marty2005subdiffusion,kou2017granular,kou2018translational}. This is also confirmed by the absence of a two-step relaxation in the self-intermediate scattering function (see Extended Data Fig.~3).

The absence of caging is related to the fact that granular particles have a rough surface which engenders to the potential energy landscape (PEL) of the system a micro-corrugation on a length scale that is much smaller than $d$, absent in atomic systems or pure hard-spheres, and sketched in Fig.~\ref{fig_1_msd}(b). This corrugation permits the particles to accumulate an irreversible displacement even at the smallest $\Gamma$ hence preventing caging. Starting from a local minimum in the PEL, the creep dynamics corresponds thus to the exploration of the local metabasin (MB), i.e., the set of configurations which are on the particle level very similar to the initial one. This motion, affected strongly by memory effects (see below), allows the system to slowly reach the boundary of the MB, i.e., the yielding point, at which many particles will have changed their neighbors. After yielding the memory is lost and the MSD becomes diffusive.

In the following we report the dynamics as a function of the accumulated strain, i.e., $\Delta \gamma= 4 \Delta n \Gamma$, which takes into account one part of the expected $\Gamma$-dependence of the dynamics. Figure~\ref{fig_1_msd}(c) shows that the diffusion coefficient $D$, obtained from the Einstein relation $\langle\delta h^2\rangle = 2D\Delta\gamma$, is basically constant for $\Gamma \lesssim 0.1$ and starts to grow sharply beyond this threshold. Also the yielding strain $\Delta\gamma_c$, locating the crossover from sub-diffusive to diffusive regime (see Extended Data Fig.~4), displays a maximum at $\Gamma = 0.1$. These observations indicate a change in the underlying dynamics as a function of $\Gamma$, directly linked to the way how the system (i) explores its MB and (ii) yields at large $\Delta n$.

Figure~\ref{fig_1_msd}(b) illustrates that for $\Gamma \lesssim 0.1$ the system will stay inside the MB during many cycles before it yields. Since the approach to the MB boundary is slow if $\Gamma \lesssim 0.1$, crossing the boundary will involve many particles at a time, i.e., this yielding is a highly collective process. In contrast to this, if $\Gamma \gtrsim 0.1$ the particles can leave the MB during each cycle, notably at maximum strain, but due to friction and particle roughness their trajectories are largely reversible. Since this reversibility is not perfect, some particles will have left the original MB and thus the MB is slowly drained during the cycling. In this case the yielding, occurring once many particles have switched their neighbors, is no longer a cooperative process since it occurs very gradually. 
In contrast to this, simulations for model glasses~\cite{fiocco2013oscillatory} or soft-sphere jammed states~\cite{kawasaki2016macroscopic,das2020unified}, in which the PEL has no micro-corrugation, found that $D$ is only non-zero above a certain threshold in $\Gamma$, thus a $\Gamma$-dependence that is very different from the one found here.

\begin{figure}
\centering
\includegraphics[width=12cm]{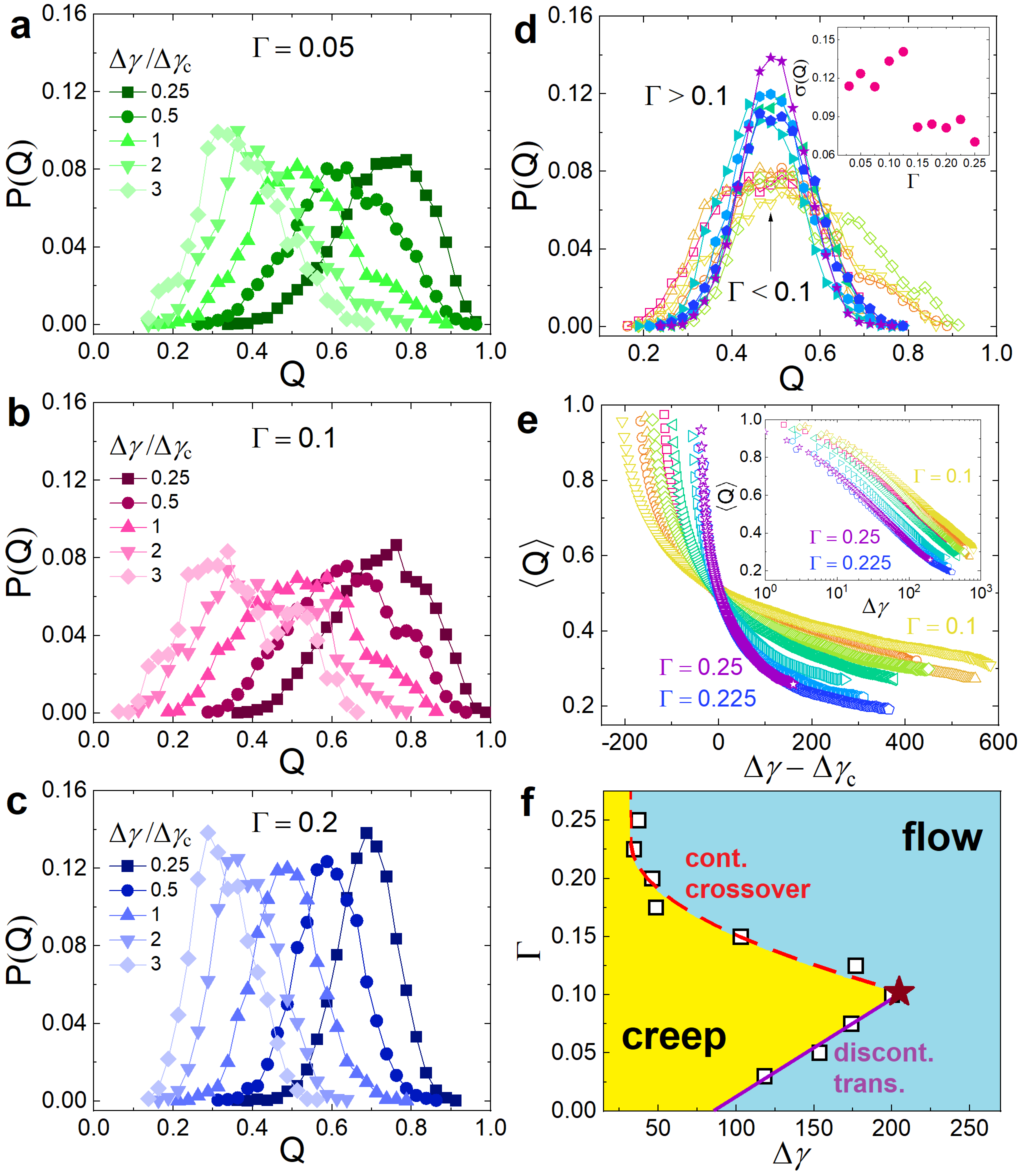}
\caption{{\bf Yielding as a phase transition.} (a-c) Overlap function distribution $P(Q)$ as a function of $\Delta\gamma/\Delta\gamma_c$ for $\Gamma = 0.05$, $0.1$, and $0.2$. $P(Q)$ shifts to smaller values as $\Delta \gamma/ \Delta \gamma_c$ grows (see legends). (d) $P(Q)$ at the yielding point $\Delta\gamma = \Delta\gamma_c$ shows the presence of two master curves. Inset: The standard deviation of $P(Q)$ versus $\Gamma$ has a sharp transition at $\Gamma \approx 0.1$. (e) $\langle Q\rangle$ as a function of $\Delta\gamma - \Delta\gamma_c$. The decay is slowest for $\Gamma \approx 0.1$ indicating the presence of a critical slowing down close to a critical point. Inset: Also $\langle Q\rangle$ versus $\Delta\gamma$ shows a slowing down at $\Gamma \approx 0.1$. (f) Dynamic phase diagram. For any $\Gamma$, the system evolves from a creep (sub-diffusion) to flow (diffusion) regime as $\Delta\gamma$ grows and yields at $\Delta \gamma =\Delta \gamma_c$. This yielding is first-order-like for $\Gamma < 0.1$ (purple solid curve) but shows only a continuous crossover for $\Gamma > 0.1$ (red dashed curve). $\Gamma = 0.1$ corresponds to the critical point (star).
In (d) and (e) the color codes are the same as in Fig.~\ref{fig_1_msd}(a).
\label{fig_4_olp}  
}
\end{figure}

The mentioned reversibility of the particle motion can be inferred directly from the smallness of the MSD at $\Delta n = 1$ in Fig.~\ref{fig_1_msd}(a), indicating that the dynamics is not Markovian but has instead a significant memory~\cite{metzler2014anomalous}. To quantify this memory we consider the correlation function between the displacements of particle $i$ at two consecutive intervals of length $\Delta \gamma$:
\begin{equation}
\label{memof}
M(\Delta\gamma) = -\langle \delta x_i(2\Delta \gamma) \delta x_i(\Delta \gamma) \rangle / \langle \delta x_i(\Delta \gamma)^2\rangle,
\end{equation}

\noindent
where $\delta x_i(t) = x_i(t) - x_i(t-\Delta \gamma)$. Figure~\ref{fig_1_msd}(d) demonstrates that $M$ is significantly positive at small $\Delta \gamma$, i.e., the displacements are anti-correlated, and $M$ decays with a $\Gamma$-dependent rate. For large $\Delta\gamma$ the memory vanishes, i.e., after yielding the dynamics becomes Markovian. We determine the strain $\Delta \gamma_M$ at which the memory is lost, $M(\Delta \gamma_M) =0$, and find that $\Delta \gamma_M(\Gamma)$ tracks $\Delta \gamma_c(\Gamma)$ very well, i.e., $\Delta \gamma_c \approx 0.4\Delta \gamma_M$, as shown in Fig.~\ref{fig_1_msd}(c). Hence the creep dynamics is accompanied by strong memory and once the initial memory has been halved the system yields.\\

{\bf Yielding as a phase transition}

The two distinct regimes in the yielding dynamics indicate that the nature of the corresponding dynamic phase transition, studied in simulations with simple shear~\cite{jaiswal2016mechanical,ozawa2018random}, changes with $\Gamma$.
One standard approach to probe the properties of a phase transition is to use as an order parameter the overlap $Q(\Delta \gamma)\in [0,1]$, which measures the similarity of two configurations separated by $\Delta \gamma$ (see Methods).
One expects that $\langle Q \rangle $ decreases with increasing $\Delta \gamma$ and the distribution  $P(Q)$ allows to identify the nature of the phase transition. 
Figures~\ref{fig_4_olp}(a-c) show that $P(Q)$ are close to $1$ for small $\Delta \gamma/\Delta\gamma_c$, i.e., most particles have not yet moved significantly. With increasing $\Delta\gamma/ \Delta\gamma_c$, $P(Q)$ shifts to the left before converging to the random distribution. For all $\Delta \gamma$ considered, the width of $P(Q)$ is significantly larger for $\Gamma \leq 0.1$ than for $\Gamma=0.2$. This is quantified in Fig.~\ref{fig_4_olp}(d) by plotting $P(Q)$ at $\Delta \gamma= \Delta \gamma_c$ for different $\Gamma$, resulting in two master curves: A wide one for $\Gamma \leq 0.1$ and a narrow one for $\Gamma >0.1$. Such a $\Gamma$-dependence is supported by the inset of Fig.~\ref{fig_4_olp}(d), which demonstrates that the standard deviation of $P(Q)$ strongly drops at $\Gamma \approx 0.1$.

\begin{figure*}[t!]
\centering
\includegraphics[width=12cm]{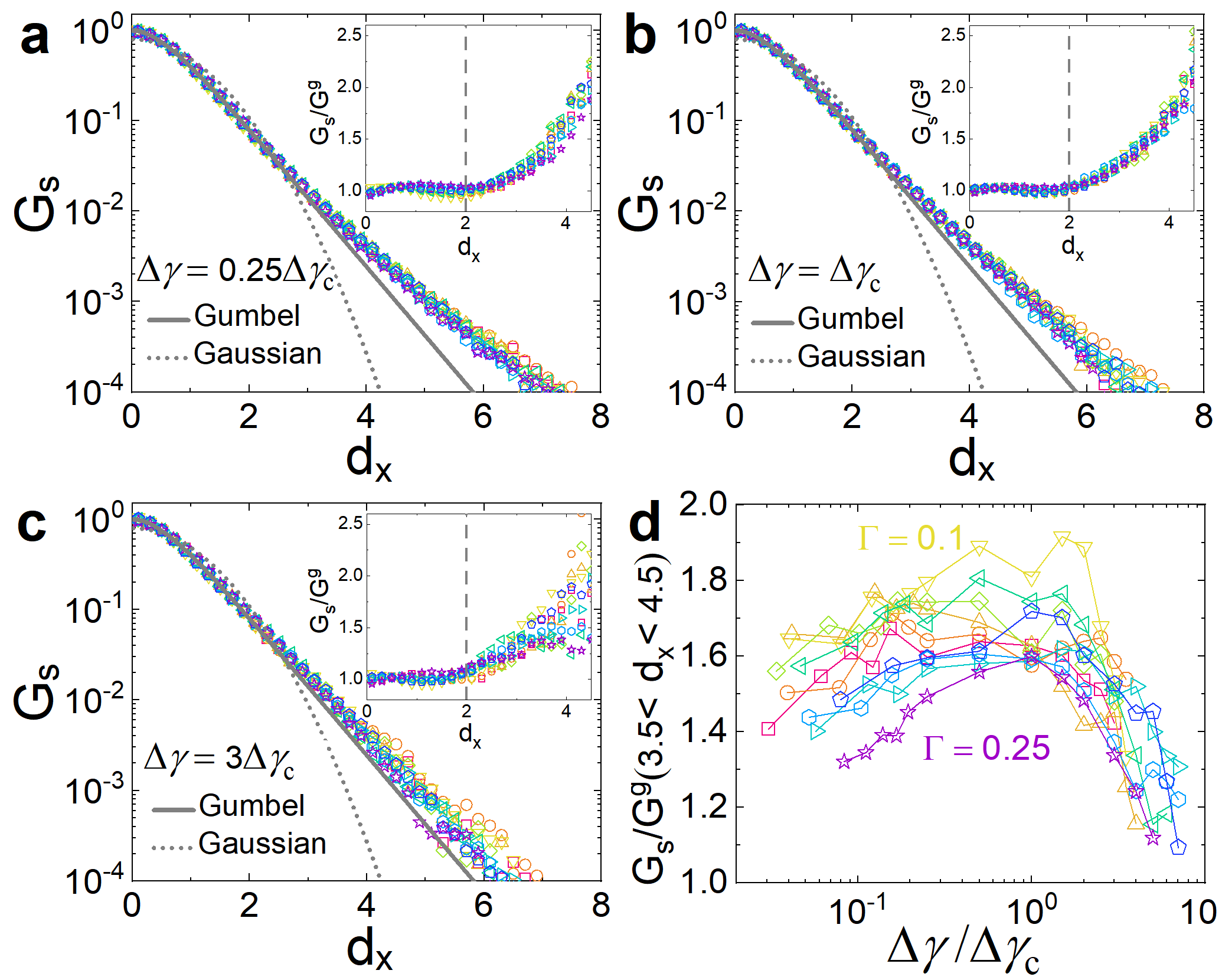}
\caption{ {\bf Self-part of the Van Hove function $G_s$ for different $\Gamma$.} The particle displacement is expressed in terms of the normalized distance $d_x = |\delta x|/\sqrt{\langle \delta x^2\rangle}$. 
The color codes are the same as in Fig.~\ref{fig_1_msd}(a).
Panels (a)-(c) correspond to $\Delta \gamma=0.25 \Delta \gamma_c$, $\Delta\gamma_c(\Gamma)$, and $3\Delta\gamma_c(\Gamma)$, respectively.
Solid and dashed curves are, respectively, the Gumbel [Eq.~(\ref{gumbel})] and Gaussian laws, showing that $G_s$ is non-Gaussian even at small $d_x$. Insets show the ratio between $G_s$ and the Gumbel law $G^g$. The vertical dashed lines at $d_x = 2$ mark the distance beyond which $G_s$ is no longer described well by the Gumbel law. For $d_x \leq 2$ the Gumbel law is universal for all $\Gamma$. (d) Intensity of the excess exponential tail obtained by averaging $G_s/G^g$ in the interval $d_x\in [3.5,4.5]$. These curves show a maximum at $\Delta \gamma \approx\Delta \gamma_c$, i.e., at the yielding point $G_s$ has the most pronounced tail.\\
\label{fig_2_dh} 
}
\end{figure*}

While the $P(Q)$ documented in Ref.~\cite{jaiswal2016mechanical} showed the double peak structure signalling a first order transition in which the two phases of the system can coexist, here we do not see clear evidence for this. However, it is well known that finite size effects do smear out such a structure, leaving only a single broad peak~\cite{jaiswal2016mechanical}, which might explain the absence of a double peak due to the moderate particle number that we consider. Alternatively one can argue that the free energy barrier between the two phases is small since the micro-corrugation of the PEL allows the existence of many locally stable particle configurations that cannot be clearly assigned to one of the two phases. In other words, the surface tension between the two phases is small, and hence the first order transition is weak. 
A weak transition hints the vicinity of a critical point at which the transition becomes second order and the dynamics shows a critical slowing down. Upon a further increase of the external parameter (here $\Gamma$), the transition ceases to exist. 
This scenario is indeed compatible with our data in that the total strain $\Delta \gamma_c$ at which the yielding occurs shows a maximum, Fig.~\ref{fig_1_msd}(c), and also the $\Delta \gamma$-dependence of the order parameter, $\langle Q \rangle (\Delta \gamma)$, shows a non-monotonic dependence on $\Gamma$, Fig.~\ref{fig_4_olp}(e). (Also here the divergence expected at the critical point is rounded off because the dynamics blurs the two phases.) For $\Gamma \geq 0.1$ the $\Delta \gamma$-dependence of $P(Q)$ indicates that there is no longer a phase transition but just a smooth crossover between the two phases, Fig.~\ref{fig_4_olp}(c). Figure~\ref{fig_4_olp}(f) summarizes the different behaviors as a function of $\Delta \gamma$ and $\Gamma$.\\

{\bf Dynamical heterogeneity} 

To obtain a microscopic understanding of the yielding, we investigate the dynamics on the particle level. Figures~\ref{fig_2_dh}(a-c) show the distributions of the particle displacements in the $x$-or $y$-direction, i.e., the self-part of the Van Hove function $G_s(\delta x,\Delta \gamma$)~\cite{binder2011glassy}, for $\Delta\gamma = 0.25\Delta\gamma_c$, $\Delta\gamma_c$, and $3\Delta\gamma_c$. Plotting $G_s$ for different $\Gamma$ as a function of the rescaled distance $d_x= |\delta x|/\sqrt{\langle \delta x^2 \rangle}$ roughly results in a master curve. In contrast to thermal systems the shape of this curve is clearly non-Gaussian even for small $\Delta \gamma$ (dotted lines), demonstrating the presence of particle motion that is faster than expected for a Gaussian process.

For $d_x < 2$ the distribution can be fitted well by a Gumbel law~\cite{kou2017granular}
\begin{equation}
\label{gumbel}
G^g(d_x) = B(\lambda)\exp[-d_x/\lambda-\exp(-d_x/\lambda)]
\end{equation}
(solid lines), where $\lambda = 0.57$ characterizes the shape of the Gumbel distribution and $B(\lambda) = 2.72$ is a normalization constant. If $d_x>2$, $G_s$ exceeds the Gumbel law and shows an exponential decay (see insets). 
The deviation of this excess tail from the Gumbel law is quantified in Fig.~\ref{fig_2_dh}(d), which presents the ratio $G_s(d_x)/G^g(d_x)$ averaged over $d_x \in [3.5,4.5]$. This deviation attains a maximum at $\Delta\gamma \approx \Delta\gamma_c$ for all $\Gamma$, signaling that at yielding the distribution is widest and hence one has maximal dynamical heterogeneity. The largest excess is found for $\Gamma \approx 0.1$, which is in line with theoretical arguments that non-Markovian processes with a significant memory give rise to a pronounced tail in the Van Hove function~\cite{Burov2022talk}. 

\begin{figure*}
\centering
\includegraphics[width=14cm]{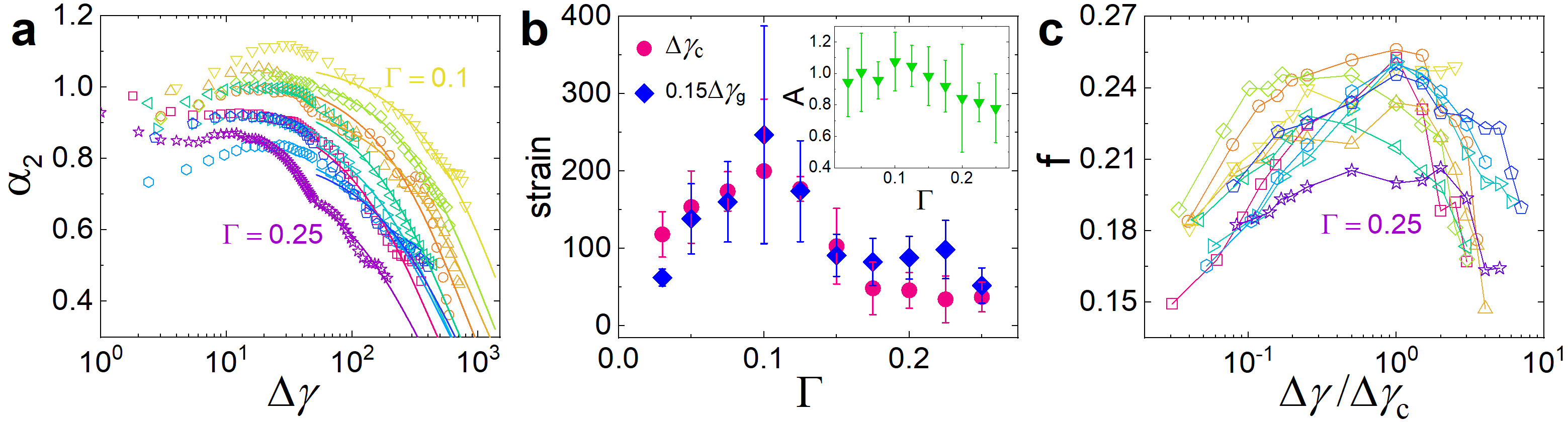}
\caption{{\bf Dynamical heterogeneity as a function of $\Delta\gamma$ and $\Gamma$.} (a) Non-Gaussian parameter $\alpha_2$ as a function of $\Delta\gamma$ for different $\Gamma$. Solid curves are exponential fits $\alpha_2 = A\cdot \exp(-\Delta\gamma/\Delta\gamma_g)$ at large $\Delta \gamma$. (b) The $\Gamma$-dependence of $\Delta \gamma_g$ tracks the one of $\Delta\gamma_c(\Gamma)$, i.e., $\Delta\gamma_c \approx 0.15\Delta\gamma_g$. Inset shows that $A(\Gamma)$ mildly peaks at $\Gamma \approx 0.1$. (c) Fraction of particles involved in the largest connected cluster (two particles are defined as connected if the center distance is smaller than $1.2$ times their average diameter), among the top $10\%$ mobile particles as a function of $\Delta\gamma/\Delta\gamma_c$. In (a) and (c) the color codes are the same as in Fig.~\ref{fig_1_msd}(a).
\label{fig_3_alpha2}  
}
\end{figure*}

We quantify the evolution of the shape of $G_s$ via the non-Gaussian parameter $\alpha_2 = \langle \delta x^4 \rangle/(3\langle \delta x^2\rangle^2) - 1$. Figure~\ref{fig_3_alpha2}(a) shows that at small $\Delta\gamma$, $\alpha_2$ varies little and is significantly larger than the value expected from the Gumbel law, $\approx 0.5$, in agreement with Fig.~\ref{fig_2_dh}(d), i.e., the exponential tail contributes considerably to $\alpha_2$. $\alpha_2$ starts to drop quickly for $\Delta\gamma \gtrsim \Delta\gamma_c$, and can be described well by an exponential, $\alpha_2(\Delta \gamma) = A(\Gamma)\exp(-\Delta \gamma/\Delta \gamma_g(\Gamma))$, where $\Delta \gamma_g$ is the strain scale for the recovery of Gaussian dynamics and $A$ is the amplitude of $\alpha_2$. Figure~\ref{fig_3_alpha2}(b) reveals that $\Delta \gamma_g(\Gamma)$ perfectly tracks $\Delta\gamma_c(\Gamma)$, similar to $\Delta\gamma_M$ in Fig.~\ref{fig_1_msd}(c), and also $A$ peaks mildly at $\Gamma \approx 0.1$ (Inset). Hence we conclude that the dynamical heterogeneities are indeed largest at around $\Gamma=0.1$.

As a direct probe of dynamical heterogeneity, we determine the spatial arrangement of the fastest particles (top $10\%$) by calculating the number of particles belonging to the largest connected cluster (defined via a nearest neighbor criterion). Figure~\ref{fig_3_alpha2}(c) presents this number, normalized by the total number of fast particles, $f$, and it peaks at $\Delta\gamma/\Delta \gamma_c \approx 1$, i.e., yielding is accompanied by a maximal cooperativity, in agreement with Fig.~\ref{fig_2_dh}(d). A random choice of 10\% of the particles in the sample gives a $f\approx 0.04$, well below the values we find here, demonstrating that the observed clustering is indeed significant. (See Extended Data Fig.~5 for the cluster size distribution.) Consistent with the phase diagram in Fig.~\ref{fig_4_olp}(f), dynamical heterogeneity peaks at yielding and $\Gamma \approx 0.1$ as displayed in Figs.~\ref{fig_2_dh} and~\ref{fig_3_alpha2}.\\

{\bf Conclusions}

Under simple shear conditions, yielding is associated with a drop in the stress-strain curve, signalling the transition from elastic to plastic behavior. 
For this type of driving, the mechanical response of granular materials will be very similar~\cite{nicolas2018deformation} since the PEL's micro-corrugation is irrelevant. 
In contrast to this, the cyclic shear considered here permits to investigate the highly non-trivial effect of this micro-corrugation and hence to detect mechanical 
features of the system inaccessible in a simple shear setup.
In view of the minimal ingredients needed to generate the micro-corrugation, we expect the surprising creep dynamics and the associated yielding behavior reported here to be generic features of granular materials and hence to be important in a multitude of situations, such as small tremors in geo-sciences or aging of civil engineering structures. 

The details of the yielding dynamics, i.e., the nature of the phase transition, will depend on the micro-roughness, shape, as well as the friction coefficient of the particles, since all these parameters influence the micro-corrugation of the PEL and hence the system dynamics. It will thus be important to study how these quantities affect the fractional diffusion behavior of granular materials, or
the strength of the memory. How this dependence can be included in the present theoretical approaches is not clear and thus remains a challenge for the future. 
We also mention that for simple shear it is custom to classify the yielding as either ductile or brittle. Our results show that for granular materials the nature of yielding depends strongly on the driving protocol, i.e., simple shear vs. cyclic shear, and also on the shear amplitude $\Gamma$. This dependence, which is absent in more standard disordered materials, suggests that for granular materials such a classification might not be possible.
Advancing on these points will lead to a fundamental understanding of granular rheology and thus permit to develop a new holistic view of the failure of complex materials.


\providecommand{\noopsort}[1]{}\providecommand{\singleletter}[1]{#1}%

\clearpage
\newpage

{\bf Methods}\\

{\bf Quasistatic shear} The cyclic shear is driven by a step motor attached to the bottom plate of the shear box. The shear rate is $\dot{\gamma} \approx 0.13$/s, giving a dimensionless inertia number $I = \dot{\gamma}d\sqrt{\rho/P} \approx 6\times 10^{-4}$, where we estimate the pressure to be given by $P \approx \rho gL$ and $L \approx 24d$. This value corresponds to a quasistatic shear condition~\cite{gdr2004dense}.

{\bf Overlap function:} For a system with $N$ particles the overlap function, which quantifies the similarity of two configurations (here separated by a strain $\Delta \gamma$), is defined as 
\begin{equation}
\label{olpfun}
  Q(\Delta \gamma) = \frac{1}{N}\sum_{i=1}^N \Theta(c - |\delta x_i(\Delta \gamma)|),
\end{equation}
where $\delta x_i$ is the particle displacement, $\Theta(\cdot)$ is the Heaviside step function, and $c$ is a preset threshold. By definition, $0 \leq Q \leq 1$, and $Q$ decreases as the system moves away from its initial configuration. In practice, we divide the cubic probe space into $2\times 2\times 2$ non-overlapping subsystems, each having $N \approx 500$, to increase the number of measurements of $Q$. Then, for given $\Gamma$ and $\Delta\gamma$, $Q$ is sampled from different subsystems, starting configurations, $x$-and $y$-directions, as well as $3\sim 5$ independent realizations. To calculate $Q$ we choose $c = 1.15\sqrt{ \langle\delta x^2(\Delta\gamma_c)\rangle}$, which makes that $\langle Q(\Delta \gamma_c)\rangle = 0.5$. This threshold must be chosen to depend on $\Gamma$, since the MSD in the subdiffusive regime changes strongly with $\Gamma$, see Fig.~\ref{fig_1_msd}(a).

{\bf Data availability.} The data that support the findings of this study are available from the corresponding authors on reasonable request.

{\bf Acknowledgments} We are grateful to S.~Burov for discussions. This work is supported by the National Natural Science Foundation of China (No. 11974240), and the Science and Technology Commission of Shanghai Municipality (No. 22YF1419900). Y. Y. acknowledges support from the fellowship of China Postdoctoral Science Foundation (No. 2021M702151). W. K. is a senior member of the Institut Universitaire de France.

{\bf Author contributions} Y.Y., Y.W. and W.K. designed the research. Y.Y., Z.Z., Y.X., H.Y. and S.Z. performed the experiment. Y.Y., W.K., and Y.W. analysed the data and wrote the paper.

{\bf Competing interests} The authors declare no competing interests.

{\bf Correspondence and requests for materials} should be addressed to Yujie Wang and Walter Kob.

\clearpage
\newpage
{\it \bf {Supplemental Material}}

\renewcommand{\figurename}{Extended Data Fig.}
\setcounter{figure}{0}
\setcounter{page}{1}

\begin{figure*}[th]
\centering
\includegraphics[width=18cm]{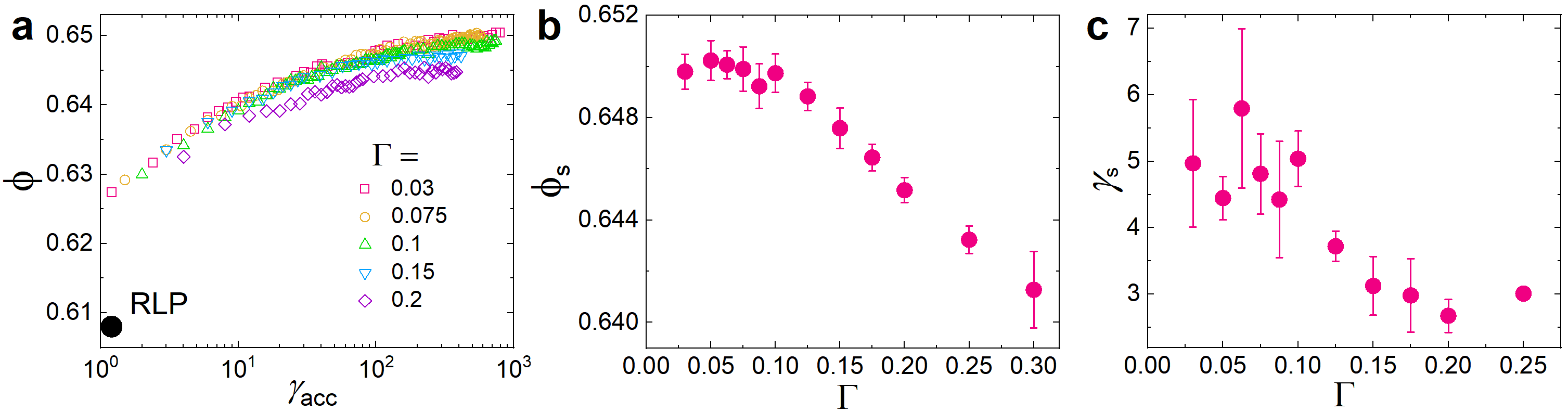}
\caption{Transient compaction process. (a) Packing fraction $\phi$ as a function of accumulated strain $\gamma_{\rm acc}$ (defined similarly to $\Delta\gamma$ but now for a dynamics that is not in steady state) for different $\Gamma$, evolving from initially deposited packings, i.e., random loose packings (RLP, marked by black dot) with $\phi_{\rm RLP} \approx 0.608$, towards the steady state. This process can be characterized by a stretched exponential law $\phi = \phi_{\rm s} + (\phi_{\rm RLP} - \phi_{\rm s})\exp[-(\gamma_{\rm acc}/\gamma_{\rm s})^{\beta}]$, where $\beta = 0.35$, $\phi_{\rm s}$ is the steady-state packing fraction, and $\gamma_{\rm s}$ is the compaction strain scale. (b) and (c) show, respectively, $\phi_s$ and $\gamma_s$ versus $\Gamma$ and one observes a crossover at $\Gamma \approx 0.1$. Error bars represent the standard deviations from different realizations.\\
}
\label{fig_S1_compact}
\end{figure*}

\begin{figure*}[th]
\centering
\includegraphics[width=12.5cm]{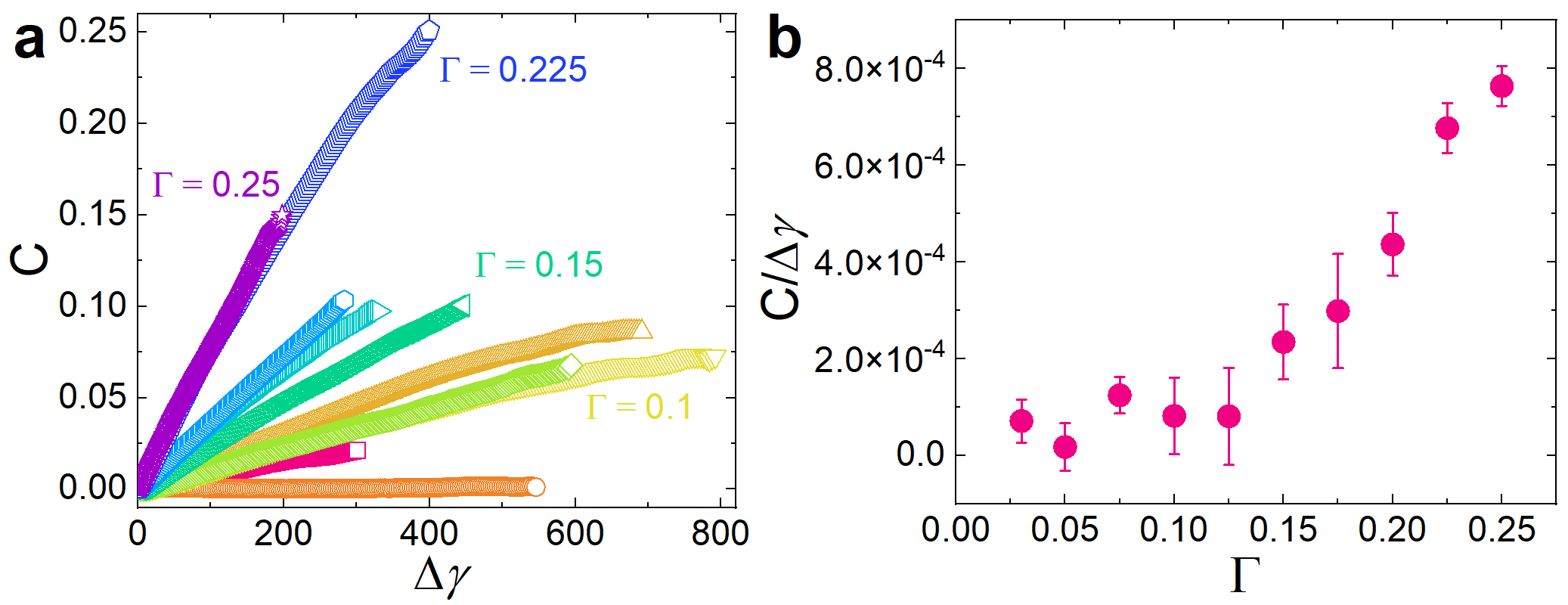}
\caption{ Convection strength. (a) Up/down asymmetry in the particle motion, characterized by $C(\Delta\gamma) = \langle z\rangle /\sqrt{\langle \delta x^2(\Delta \gamma)\rangle}$, as a function of $\Delta\gamma$.  $C$ grows linearly with $\Delta\gamma$ with slopes that depend on $\Gamma$, indicating an upward motion, i.e., convection. (b) The associated slope $C/\Delta\gamma$ as a function of $\Gamma$ is basically zero for $\Gamma \lesssim 0.1$ and increases significantly for larger $\Gamma$. This is further evidence that the dynamics changes at $\Gamma \approx 0.1$ allowing the convection to set in for large $\Gamma$. Error bars represent the standard deviations from different realizations.}
\label{fig_S2_convect}
\end{figure*}

\begin{figure*}[th]
\centering
\includegraphics[width=18cm]{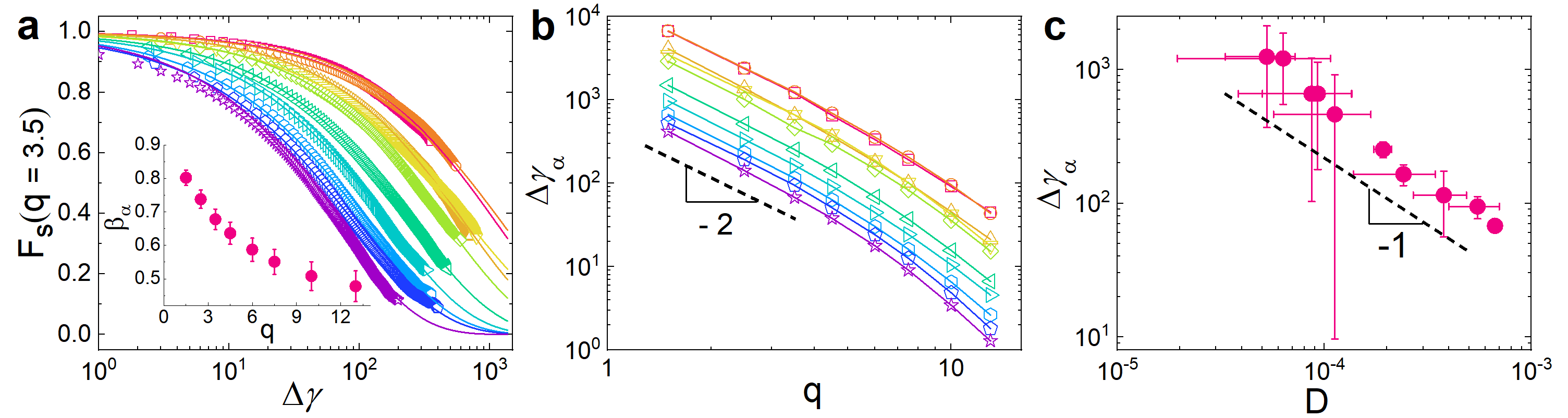}
\caption{Self-intermediate scattering function $F_s(q,\Delta \gamma)$. (a) $F_s(q,\Delta \gamma)$ for $q = 3.5$, i.e., the first peak in static structure factor, and different $\Gamma$. Solid curves indicate a fit with a stretched exponential $F_s(q,\Delta \gamma) = \exp[-(\Delta\gamma/\Delta\gamma_{\alpha})^{\beta_{\alpha}}]$, where $\Delta\gamma_{\alpha}$ is the relaxation strain scale. Inset: Stretched exponent $\beta_{\alpha}$ as a function of $q$. (b) $\Delta\gamma_{\alpha}$ as a function of $q$ for different $\Gamma$. The scaling $\Delta\gamma_{\alpha} \propto q^{-2}$ for small $q$ (dashed line) indicates the Gaussian dynamics for large length scale, in agreement with the limit $\beta_{\alpha} \rightarrow 1$ for $q \rightarrow 0$ shown in the inset of panel (a). The colors of the curves in (a) and (b) are the same as in Fig.~1(a) of the main text. (c) For $q=3.5$ one finds that $\Delta\gamma_{\alpha} \propto D^{-1}$ for different $\Gamma$ (dashed line), indicating that $F_s(q,\Delta \gamma)$ conveys the same information as the MSD presented in the main text. Error bars represent the standard deviations from different realizations.}
\label{fig_S3_isf}
\end{figure*}

\begin{figure*}[th]
\centering
\includegraphics[width=12cm]{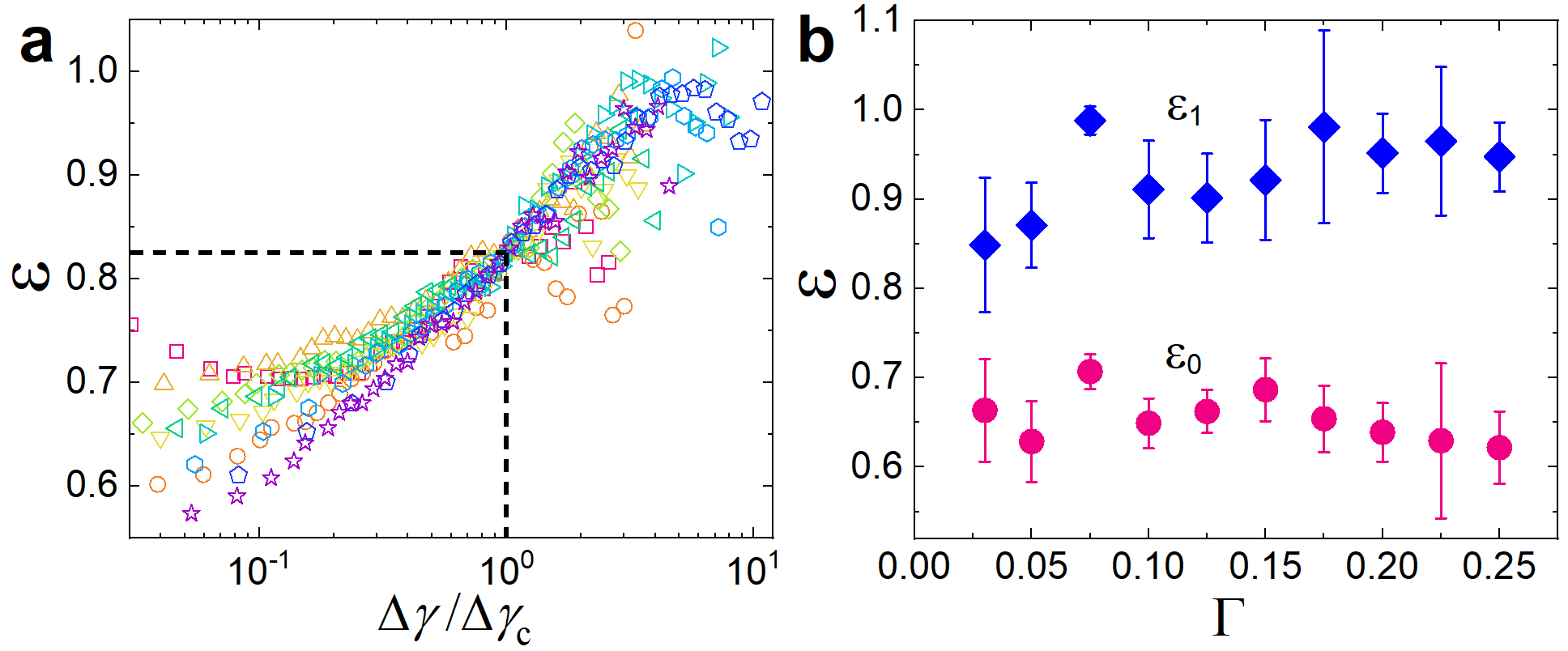}
\caption{ Scaling of the MSD to determine the yielding strain $\Delta \gamma_c$. (a) Power-law exponent $\varepsilon = d\log(\langle \delta h^2\rangle)/d\log(\Delta\gamma)$ as a function of $\Delta\gamma / \Delta\gamma_c$. The yielding strain $\Delta\gamma_c$ is located at $\varepsilon = 0.825$ (vertical and horizontal dashed lines). The evolution of $\varepsilon$ can be described by  $\varepsilon = \varepsilon_1 + (\varepsilon_0 - \varepsilon_1)\exp(-\Delta\gamma / \Delta\gamma_c)$, where $\varepsilon_0$ and $\varepsilon_1$ are fit parameters. (b) $\Gamma$-dependence of the parameters $\varepsilon_0$ and $\varepsilon_1$. This graph confirms that the two exponents in the MSD are independent of $\Gamma$. Error bars represent the standard deviations from different realizations.
} 
\label{fig_S4_msd_slope}
\end{figure*}

\begin{figure*}[th]
\centering
\includegraphics[width=12cm]{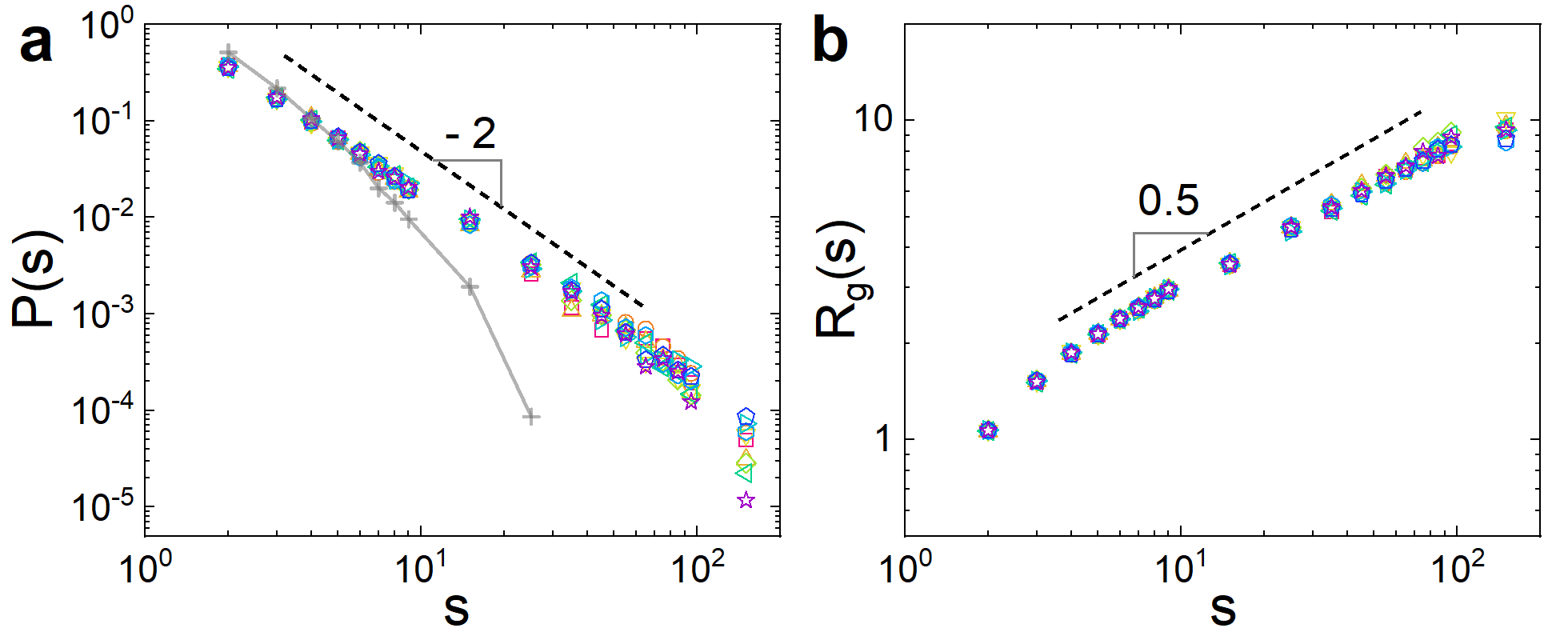}
\caption{ Characterizing the clusters of top $10\%$ fastest particles, i.e.,~the same threshold used in Fig.~\ref{fig_3_alpha2}(c). Color of symbols are the same as in Fig.~1(a) of the main text. (a) Cluster size distribution $P(s)$ for different $\Gamma$ at the yielding point, i.e., $\Delta\gamma = \Delta\gamma_c$. An approximately universal scaling $P(s) \propto s^{-2}$ (dashed line) is found. $P(s)$ from the randomly chosen $10\%$ (crosses with line) deviates strongly from this master curve. (b) The average gyration radius $R_g(s)$ as a function of $s$ also shows a universal scaling $R_g(s) \propto s^{0.5}$ (dashed line). 
} 
\label{fig_S5_clust}
\end{figure*}

\end{document}